\documentclass[12pt]{article}

\global\arraycolsep=1pt
\usepackage{graphicx}
\usepackage{amssymb}
\renewcommand{\thefootnote}{\fnsymbol{footnote}}

\newcommand{\beq}{\begin{equation}}
\newcommand{\eeq}{\end{equation}}
\newcommand{\beqa}{\begin{eqnarray}}
\newcommand{\eeqa}{\end{eqnarray}}

\begin{document}

%
%
\begin{titlepage}

\begin{flushright}
       \normalsize
       OCU-PHYS 247  \\
       hep-th/0605129 \\
\end{flushright}

\vspace{5mm}

\begin{center}
{\Large \bf  Explicit Toric Metric on \\
Resolved Calabi-Yau Cone}
\end{center}

\vspace{3mm}

\begin{center}
{Takeshi Oota$^1$\footnote{e-mail: \texttt{toota@sci.osaka-cu.ac.jp}}
and 
Yukinori Yasui$^2$\footnote{e-mail: \texttt{yasui@sci.osaka-cu.ac.jp}}}
\end{center}

\begin{center}
$^1${\sl Osaka City University, 
Advanced Mathematical Institute (OCAMI) \\
3-3-138 Sugimoto, Sumiyoshi, Osaka 558-8585, Japan}\\
\vspace{4mm}
$^2${\sl Department of Mathematics and Physics,
Graduate School of Science, \\
Osaka City University\\
3-3-138 Sugimoto, Sumiyoshi, Osaka 558-8585, Japan}\\
\end{center}


\begin{abstract}
We present an explicit non-singular complete toric Calabi-Yau metric
using the local solution recently found by Chen, L\"{u}
and Pope. This metric gives a new supergravity solution representing 
D3-branes.
\end{abstract}

\vfill

\end{titlepage}

\renewcommand{\thefootnote}{\arabic{footnote}}
\setcounter{footnote}{0}


D3-branes on the tip of toric Calabi-Yau cones have been extensively
studied in connection with the AdS/CFT correspondence \cite{ADSCFT}.
It is natural to consider the deformations of cone metrics
in order to explore non-conformal theories \cite{PP,PT0,PT,pal,SZ,BMPT}.

In this letter, we study a Calabi-Yau metric,
~i.e. Ricci-flat K\"ahler metric,
constructed as the BPS limit of the six dimensional Euclideanised
Kerr-NUT-AdS black hole metric \cite{CLP}. In the black hole with
equal angular momenta, 
the corresponding Calabi-Yau metric is of the form
\begin{eqnarray}
\label{CYM}
g&=&\frac{(1-x)(1-z)}{3}(d\theta^2+\sin^2 \theta d\phi^2)
+\frac{(1-x)(x-z)}{f(x)}dx^2+\frac{(1-z)(z-x)}{h(z)}dz^2 \\
&+&\frac{f(x)}{9(1-x)(x-z)}(d\psi-\cos\theta d\phi+
z(d\beta+\cos \theta d\phi))^2\nonumber \\
&+&\frac{h(z)}{9(1-z)(z-x)}(d\psi-\cos\theta d\phi+
x(d\beta+\cos \theta d\phi))^2,\nonumber 
\end{eqnarray}
where
\beq
f(x)=2 x^3-3 x^2+a,~~h(z)=2 z^3-3 z^2+b
\eeq
with two parameters $a$ and $b$.
We assume that the roots $x_i$ ($i=1,2,3$) of $f(x)=0$
are all distinct and real,
and further they are ordered as $z_1 < x_1 < x_2 < x_3$ for the 
smallest real root $z_1$ of $h(z)=0$. 
In order to avoid a curvature singularity
we take the coordinates $x$ and $z$
to lie in the region $z \leq z_1 < x_1 \leq x \leq x_2 < 1$.
Indeed, it is easy to see that such a singularity appears at $z=x$. 

For $r \to \infty~(z =-r^2/2)$ , the metric tends to a cone metric
$ dr^2+r^2 \bar{g}$,
where
\begin{eqnarray}
\bar{g}&=&\frac{1-x}{6}(d\theta^2+\sin^2\theta d\phi^2)+\frac{1-x}{2 f(x)}
dx^2+\frac{f(x)}{18(1-x)}(d\beta+\cos \theta d\phi)^2 \\
&+&\frac{1}{9}
(d\psi-\cos\theta d\phi+x(d\beta+\cos \theta d\phi))^2.\nonumber
\end{eqnarray}
This metric yields the Sasaki-Einstein metric $Y^{p,q}$ when
we impose a suitable condition for the parameter $a$ \cite{GMSW2}.

Next let us look at the geometry near $z=z_1$. We introduce
new coordinates given by
\beq
\varphi_{1}=-\frac{1}{2}(\psi-z_1 \beta),~~
\varphi_{2}=\frac{1}{1-z_1}(\psi+z_1 \beta),~~
\varphi_{3}=\phi.
\eeq
Then the metric behaves as
\beq
g \simeq du^2+u^2 d\varphi_{1}^2+g^{(4)}, 
\eeq
where $u^2=2(z_1-x)(z_1-z)/(3z_1)$. Therefore,
the periodicity of $\varphi_1$
should be $2 \pi$ in order to avoid
an orbifold singularity.
The four dimensional K\"ahler metric $g^{(4)}$
is given by
\begin{eqnarray}
\label{4DK}
g^{(4)}&=&\frac{(1-x)(1-z_1)}{3}
(d\theta^2+\sin^2 \theta d\varphi_{3}^2)
+\frac{(1-x)(x-z_1)}{f(x)}dx^2 \\
&+&\frac{(1-z_1)^2f(x)}{9(1-x)(x-z_1)}
(d\varphi_{2}-\cos \theta d \varphi_3)^2. \nonumber
\end{eqnarray}
We now argue that by taking the special parameters
$a = (1/2) - (1/32)\sqrt{13}$ which 
corresponds to $Y^{2,1}$,
and $b = (1/4)(137 + 37 \sqrt{13} )$,
the four dimensional space with metric $g^{(4)}$
is a non-trivial $S^2$-bundle over $S^2$, i.e. 
the  first del Pezzo surface $dP_1$.
To see this,  introduce a radial coordinate 
$y^2=2(x_i-z_1)\mid x-x_i \mid/\mid 3 x_i \mid $ on  
the $(x,\varphi_2)$-fibre space defined by
fixing the $S^2$ coordinates
$\theta$ and $\varphi_3$ in (\ref{4DK}). Then, the fibre metric
 near boundary $x=x_i$ $(i=1,2)$ is written as 
\beq
dy^2+\left(\frac{x_i(1-z_1)}{x_i-z_1} \right)^2 y^2 d\varphi_2^2.
\eeq 
Using the values of $a$ and $b$, we have 
\beq
\label{x12z1}
x_1=\frac{1}{8}(1-\sqrt{13}),~~x_2=\frac{1}{8}(7-\sqrt{13}),~~
z_1=-\frac{1}{2}(2+\sqrt{13}),
\eeq
and then $x_i(1-z_1)/(x_i-z_1)= \pm 1/2$.
The apparent singularities at $x=x_i$ can be avoided by
choosing the periodicity of $\varphi_2$ to be $4 \pi$.
Thus, the $(x,\varphi_2)$-fibre space is topologically $S^2$.
On the other hand, fixing the coordinate $x$ in (\ref{4DK}), we obtain
a principal $U(1)$-bundle over $S^2$ with the Chern number
\beq
 \frac{1}{4\pi} \int_{S^2}d(-\cos \theta d\varphi_3)=1.
\eeq
The metric $g^{(4)}$ can be regarded as a metric on the
associated $S^2$-bundle of the principal $U(1)$-bundle.
The associated bundle is non-trivial since 
the Chern number is odd, and hence the total space is the $dP_1$.

Let us describe the Calabi-Yau metric (\ref{CYM}) from the point of view
of toric geometry. The metric has an isometry $T^3$, locally
generated by the Killing vector fields, $\partial/\partial \psi,~
\partial/\partial \phi$ and $\partial/\partial \beta$.
The symplectic (K\"ahler) form $\omega$ is given by
\beq
\omega =\frac{1}{3} \Bigl( d\psi \wedge d(x+z)
+d\phi \wedge d((1-x)(1-z)\cos \theta)
+d\beta \wedge d(x z) \Bigr).
\eeq
Using the following generators of the $T^3$ action \cite{MS},
\begin{eqnarray}
\frac{\partial}{\partial \phi_1} &=& \frac{\partial}{\partial \psi}
+\frac{\partial}{\partial \phi}-
\frac{\partial}{\partial \beta},\\
\frac{\partial}{\partial \phi_2} &=& \frac{\partial}{\partial \phi}
+3 \ell \frac{\partial}{\partial \beta}, \\
\frac{\partial}{\partial \phi_3} &=& -6 \ell 
\frac{\partial}{\partial \beta}, 
\end{eqnarray}
one has Darboux coordinates $(\xi^1,~\xi^2,~\xi^3)$ on which 
the symplectic form
takes the standard form $\omega= d\xi^i \wedge d \phi_i$:
\begin{eqnarray}
\xi^1 &=& \frac{1}{3}(1-x)(1-z)(1-\cos \theta),\\
\xi^2 &=& -\frac{\ell}{2}  (2 x z + 1 )
-\frac{1}{3}(1-x)(1-z)\cos \theta, \\
\xi^3 &=& \ell ( 2 x z + 1 )
\end{eqnarray}
with
$\ell^{-1} = -5+2\sqrt{13}$. 
For the range of variables: 
$0 \leq \theta \leq \pi, x_1 \leq x \leq x_2,~
z \leq z_1$, where $x_1, x_2$ and $z_1$ are given by (\ref{x12z1}), we find
the Delzant polytope (see Fig.1)
\beq
P=\{\xi=(\xi^1,\xi^2,\xi^3) \in \mathbb{R}^3 \mid 
(\xi, v_a)\geq \lambda_a, \, a=1,2,\dots, 5 \}. 
\eeq%
\begin{figure}[ht]  
  \begin{center}
  \includegraphics[height=4cm]{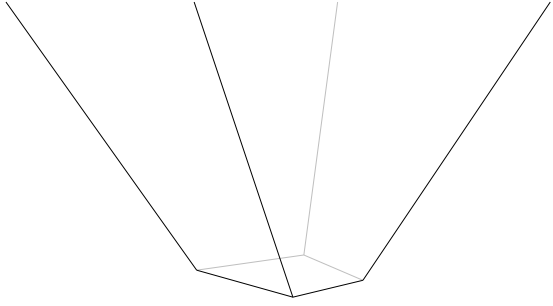} 
  \end{center}
  \caption{Delzant polytope $P$} 
\end{figure}%
\noindent
Here, each $v_a$ is a primitive element of the lattice
$ \mathbb{Z}^3 \subset \mathbb{R}^3 $ and an inward-pointing
normal vector to the two dimensional face of $P$. Explicitly,
the set of five vectors $v_a = (1, w_a)$ can be chosen as
(see Fig. 2)
\beq
w_1 = (-1,-2), \ \ 
w_2 = (0, 0), \ \ 
w_3 = (-1,0), \ \ 
w_4 = (-2, -1), \ \
w_5 = (-1,-1).
\eeq%
\begin{figure}[ht] 
  \begin{center}
  \includegraphics[height=5cm]{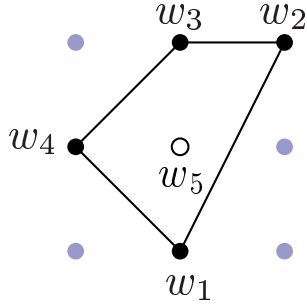} 
  \end{center}
  \caption{Toric diagram} 
\end{figure}%
\noindent
The constants $\lambda_a$ are given by
\beq
\lambda_1 = \frac{1}{72} ( 1 - 5 \sqrt{13} ), \qquad
\lambda_3 = \frac{1}{216} (29 + 17 \sqrt{13} ), \qquad
\lambda_5 = \frac{1}{54} (31 + 7 \sqrt{13} ),
\eeq
and $\lambda_2 = \lambda_4 = 0$.
The inner products $(\xi, v_a)$ are evaluated as
\begin{eqnarray}
( \xi, v_1 )  &=& \frac{2}{9} ( 1 + \sqrt{13} )
( x - x_1) ( x_1 - z )+ \lambda_1 , \\
( \xi, v_2 ) &=&
\frac{1}{3} ( 1 - x ) ( 1 - z ) ( 1 - \cos \theta), 
\nonumber \\
( \xi, v_3 ) &=& \frac{2}{27} ( 7 + \sqrt{13} )
( x_2 - x )( x_2 - z)+\lambda_3 , \nonumber \\
( \xi, v_4 ) &=&
\frac{1}{3} ( 1 - x )( 1 - z) ( 1 + \cos \theta), \nonumber \\
( \xi, v_5 ) &=& \frac{2}{27}(-2 + \sqrt{13} )
( x - z_1 )( z_1 - z)+\lambda_5 . \nonumber
\end{eqnarray}
Thus, we see that the five faces $
F_a=\{\xi \in \mathbb{R}^3 \mid 
(\xi, v_a)= \lambda_a \}$ correspond to degeneration surfaces at
$x=x_1,~\theta=0,~x=x_2,~\theta=\pi$ and $z=z_1$, respectively.

Finally, we note that
a D3-brane solution can be constructed 
from the Calabi-Yau metric given by $g$ (\ref{CYM})
with the special parameters $a$, $b$:
\begin{eqnarray}
g^{(10)} &=& H^{-1/2} g^{(3+1)}
+ H^{1/2} g, \\
F_5 &=& ( 1 + *_{10} ) \mathrm{vol}_{(3+1)} \wedge d H^{-1}. \nonumber
\end{eqnarray}
We find the warp factor $H$ as a harmonic function 
$\bigtriangleup_g H=0$~;
\begin{eqnarray}
H(z) &=& - \frac{(7-2\sqrt{13}) L^4}{27}
\log \left( \frac{8z^3-12z^2+137+37\sqrt{13}}{(2z+2+\sqrt{13})^3}
\right) \\
& &
\ \ \ -  \frac{(10-2\sqrt{13})L^4}{27\sqrt{6+2\sqrt{13}}}
\left( \frac{\pi}{2}
- \mathrm{arctan}\left(
\frac{-4z + 5 + \sqrt{13}}{3 \sqrt{6+2\sqrt{13}}}
\right)\right),
\end{eqnarray}
where the constant $L$ is given by
\beq
L^4 = \frac{4\pi^4 g_s (\alpha')^2 N}{\mathrm{Vol}(Y^{2,1})}
= 16\bigl( - 46 + 13 \sqrt{13} \bigr) \pi g_s (\alpha')^2 N.
\eeq
For large $-z = r^2/2$, the warp factor behaves as
\beq
H = \frac{L^4}{r^4} + O(1/r^6),
\eeq
while near $z = z_1$
\beq
H \simeq \frac{2(7-2 \sqrt{13}) L^4}{27} \log( z_1 - z ).
\eeq

\vspace{10mm}
\noindent
{\bf \Large Acknowledgements}

\vspace{5mm}

We would like to thank K. Maruyoshi
for useful discussions.
This work is supported by the 21 COE program 
``Constitution of wide-angle mathematical basis focused on knots.''
The work of Y.Y. is supported by 
the Grant-in Aid for Scientific Research
(No. 17540262 and No. 17540091) from Japan Ministry of Education.
The work of T.O. is supported in part by the Grant-in-Aid
for Scientific Research (No.18540285)
from the Ministry of Education, Science and Culture, Japan.




\end{document}